\documentclass{svproc}
\usepackage{url}
\usepackage{float}
\usepackage{epsfig,amsmath,amsfonts}

\begin{document}
\mainmatter              % start of a contribution
\title{Matrix solitons solutions of 
the\\ modified Korteweg-de Vries equation}
%
%\titlerunning{%KdV \& 
%mKdV matrix soliton solutions}  % abbreviated title (for running head)
\titlerunning{Matrix solitons solutions of 
the mKdV equation}
\author{Sandra Carillo\inst{1,2} \and Mauro Lo  Schiavo\inst{3}
 \and
Cornelia Schiebold\inst{4,5}}
\authorrunning{Sandra Carillo, Mauro Lo  Schiavo,  and Cornelia Schiebold} % abbreviated author list (for running head)
\tocauthor{Sandra Carillo, Mauro Lo  Schiavo,  and Cornelia Schiebold}
\institute{Universit\`a di Roma  \textsc{La Sapienza}, Dip.S.B.A.I., 16 Via A. Scarpa,  Rome, Italy,\\
\email{Sandra.Carillo@uniroma1.it},\\ WWW home page:
\texttt{https://www.sbai.uniroma1.it/{\textasciitilde{}}sandra.carillo/index.html}
\and
I.N.F.N. - Sezione Roma1,
Gr. IV - M.M.N.L.P.,  Rome, Italy\\
\and
Universit\`a di Roma  \textsc{La Sapienza}, Dip.D.I.S.G., 18 Via Eudossiana,  Rome, Italy,\\
\and
Department of Mathematics and Science Education, 
    Mid Sweden University,  Sundsvall, Sweden \\
\and
Instytut Matematyki, Uniwersytet Jana Kochanowskiego w Kielcach, Poland}

\maketitle              % typeset the title of the contribution

\begin{abstract}
Nonlinear non-Abelian  Korteweg-de Vries (KdV) and modified Korteweg-de Vries (mKdV) 
equations and their links via  B\"acklund transformations are considered. The focus  is on 
the construction of {\it soliton solutions} 
admitted by  matrix modified Korteweg-de Vries equation. Matrix equations can be viewed as 
a specialisation 
of operator equations in  the  finite dimensional case  when operators  
%are finite dimensional and, hence, 
admit a matrix representation. B\"acklund transformations  allow to reveal structural properties  %[S.Carillo, C.Schiebold, JMP 50, 2009, 073510]
\cite{Carillo:Schiebold:JMP2009} enjoyed by non-commutative KdV-type equations, 
such as the existence of a recursion operator. 
{\it Operator methods} combined with B\"acklund transformations, 
allow to construct explicit solution formulae 
%[S.Carillo, C.Schiebold, JMP 52, 2011, 053507]. %
\cite{Carillo:Schiebold:JMP2011}. 
The latter are adapted to obtain solutions admitted by the  $2\times2$ and 
$3\times3$ matrix mKdV equation. Some of these matrix solutions are visualised to 
show the {\it solitonic} behaviour  they exhibit.
A further key tool used to obtain the presented results is an {\it ad hoc} construction of computer algebra routines to implement non-commutative computations.
\end{abstract}
\section{Introduction}
So called {\it soliton equations}  are widely investigated since the late 1960's when the first exact solution of the 
Korteweg-de Vries equation was obtained, as reported, for instance, in the well known book by Calogero and Degasperis 
\cite{CalogeroDegasperis}.
 The name  {\it soliton equations} is generally used to indicate nonlinear evolution equations 
which admit exact solutions in the Schwartz space of smooth {\it rapidly decreasing functions}\footnote{It is generally assumed that $M$ is   space of functions  $u(x,t)$ which, $\forall$  fixed $t$, belong to the Schwartz space $S$ of {\it rapidly decreasing functions} on ${{\mbox{R\hskip-0.9em{}I \ }}}^n$,  in the  i.e. 
$S({{\mbox{R\hskip-0.9em{}I \ }}}^n):=\{ f\in C^\infty({{\mbox{R\hskip-0.9em{}I \ }}}^n)\!\! : \!\!\vert\!\vert f \vert\!\vert_{\alpha,\beta} < \infty, \forall \alpha,\beta\}$, where 
$\vert\!\vert f \vert\!\vert_{\alpha,\beta}:= sup_{x\in{{\mbox{R\hskip-0.9em{}I \ }}}^n} \left\vert x^\alpha D^\beta f(x)\right\vert $, $D^\beta:\!\!=\!\!\partial^\beta /{\partial x}^\beta$. %In this Section $n=1$.
}. The relevance of B\"acklund transformations  in 
studying nonlinear evolution equations is well known both under the viewpoint of finding solutions to given 
initial boundary value problems, see \cite{RogersAmes} - \cite{RogersShadwick} 
as well as in giving insight in the study of their structural properties, such as
symmetry properties, admitted conserved quantities and Hamiltonian structure, see e.g.
 \cite{JMP2018} and references therein. %,  is well established. 
The important role played by B\"acklund transformations  to investigate properties enjoyed 
by nonlinear evolution equations 
 is based on the fact that most of the properties of interest are preserved under B\"acklund transformations \cite{FokasFuchssteiner:1981},  \cite{Fuchssteiner1979}.  Thus, the construction of 
a  net of B\"acklund transformation to  connect nonlinear evolution equations  allowed 
to prove new results both in the  case of {\it scalar} equations  \cite{ActaAM2012}-\cite{BS1}, \cite{PhysA89}-\cite{PhysA90}, %\cite{IMACS2017}, \cite{ActaAM2012}, 
% \cite{BS1}, \cite{PhysA89},\cite{PhysA90}, \cite{IMACS2017}, \cite{ActaAM2012}, % \cite{RogersCarillo}
as well as in the generalised case of operator equations \cite{SIGMA2016}-\cite{JNMP2012} 
\footnote{An overview on non-commutative equations is given in \cite{Kupershmidt}.}
A comparison between the scalar (Abelian)  and the operator (non-Abelian) cases referring to 
third order KdV-type equations, B\"acklund transformations connecting them  and related 
properties is comprised in \cite{waves2-IMACS2017}. %
Notably, B\"acklund transformations indicate a way to construct 
%Under the applicative viewpoint, 
solutions to nonlinear evolution equations  \cite{JPhysA1992} 
%, on application of results in \cite{PhysA89}, 
and also to nonlinear ordinary differential equations,  \cite{2017arXiv171104304C} and \cite{WASCOM2017}. Then, the operator approach  \cite{AdenCarl}, \cite{Carl:Schiebold}, \cite{Marchenko},  \cite{Schiebold 0},  \cite{Schiebold1}, 
% combined with B\"acklund transformations
%, according to \cite{Carillo:Schiebold:JMP2011}, 
allows to construct solutions admitted by the whole hierarchies of nonlinear operator equations which are connected via B\"acklund transformations, see  
\cite{ActaAM2012}-\cite{Carillo:Schiebold:JMP2011} and references therein. 
 The special case under investigation concerns solutions of the $2\times2$ and 
 $3\times3$ mKdV matrix equation.  Matrix equations are studied in \cite{LeviRagniscoBruschi} 
 and in  \cite{Goncharenko}, where, 
respectively, solutions admitted by Burgers and KdV equations are obtained. 
A motivation for investigations on matrix  equations is connected to quantum mechanics.

The material is organised as follows.  Section \ref{1}  briefly reminds the needed notions on 
B\"acklund transformation,  recalling also the  links among  KdV-type  equations. 
In the subsequent Section \ref{3}, a short overview is provided on the method  to obtain 
matrix solutions in the  finite dimensional case. 
 Then, some  solutions of the matrix mKdV equation are graphically represented and, finally,  
 relevant  remarks and research perspectives are in the closing Section  \ref{4}.

\vskip-1em\section{Noncommutative  potential KdV,  KdV and modified KdV  hierarchies}\label{1}
The  well known potential Korteweg deVries (pKdV),  Korteweg deVries (KdV) and modified Korteweg deVries (mKdV) equations, in turn
\begin{equation}\label{kdv}
w_t= w_{xxx} + 3 w_x^2
\qquad  \textstyle{,} \qquad
u_t \ =\  u_{xxx} + 6 uu_x
\qquad  \text{and} \qquad
v_t \ =\  v_{xxx} + 6 v^2 v_x
\end{equation}
are nonlinear evolution equations in the unknown real functions, respectively 
$w$, $u$ and $v$.  
All the equations \eqref{kdv}  admit {\it soliton solutions}, see, among a wide literature, 
e.g. \cite{CalogeroDegasperis}, namely 
solutions which represent  a nonlinear waves which propagate preserving energy and shape. 
The  aim of the present investigation is to consider a generalisation of equations \eqref{kdv}, on introduction of operator valued equations according to the approach in \cite{AdenCarl},  \cite{Carl:Schiebold}. Thus, the non-commutative equations, counterpart, respectively, of  \eqref{kdv}, are given by:
\begin{equation}\label{nckdv}
W_t= W_{xxx} + 3 W_x^2,~~ U_t \ =\  U_{xxx} + 3  \{ U, U_x\},~
~\text{and}~~
V_t = V_{xxx} + 3  \{ V^2,V_{x}\},
\end{equation}
where the square and curly brackets  denote, in turn,  the commutator and the 
anti-commutator, that is $\forall T, S$,
$ [T,S]:= TS-ST$  and $ \{T,S\}:= TS+ST$.
 
According to the definition,  \cite{FokasFuchssteiner:1981}, \cite{Fuchssteiner1979}, 
 two different evolution equations, e.g. \eqref{kdv}$_1$ and  \eqref{kdv}$_2$, are termed  {\it connected via the B\"acklund transformation }
$B {(u, v)} = 0$
whenever  given two solutions they admit,   say, $u(x,t)$ and $v(x,t)$,   if
\begin{equation}
B (u(x,t),v(x,t) )\vert_{t=0} = 0 
\quad ~\text{implies}\quad  
B (u(x,t),v(x,t) )\vert_{t=\tau} = 0, \forall  \tau >0.
\end{equation}
Well known examples of B\"acklund transformations are the introduction of a 
{\it bona fide} potential 
 and the Miura transformation which, respectively,
 relate the pKdV to the KdV and the latter to the mKdV equation. 
 The non-commutative  extensions are %\vskip-1.5em
{\begin{equation}\label{Mnc}
B_1:  U-W_x=0 ~~~\hphantom{x} \hskip2em M: ~~U - i V_{x} -  V^{2}= 0~~.
\end{equation}} 
Notably, \cite{Carillo:Schiebold:JMP2009}, \cite{Schiebold2010}, the links via B\"acklund transformations, combined with the knowledge of the hereditary recursion operator of the KdV equation, allow  to construct the recursion operator admitted by the pKdV as well as by mKdV equation and, hence,  to  extend the same links to the whole corresponding hierarchies (see \cite{SIGMA2016} and references therein for details). 

        That is, given the KdV recursion operator
  $    %  \begin{equation}\label{kdv-recop} %{kdv-recop}
            \Phi(U)$ % = 
         then,  the mKdV recursion operator 
        and the pKdV recursion operator 
        are obtained \cite{Carillo:Schiebold:JMP2009}. %Hence, 
        All the corresponding members in the three  {\it hierarchies} of pKdV, KdV and mKdV equations follows to be linked via the  transformations 
        \eqref{Mnc} 
         Remarkably,  these connections \cite{Carillo:Schiebold:JMP2009} allow, given a 
        solution of the noncommutative pKdV equation,  to construct the corresponding 
        solutions of the   noncommutative KdV and mKdV equations. 

\section{Matrix soliton solutions}\label{3}
This section aims to summarise the essential steps in the construction of solutions on application of the operator method  devised in
 \cite{AdenCarl}, \cite{Carl:Schiebold},  further developed in \cite{Schiebold1},  \cite{Schiebold2}, and extended to hierarchies in
\cite{Carillo:Schiebold:JMP2011}.  A very short outline on how this method can be adopted to construct solutions of the matrix mKdV equation
 is provided. Indeed, the study in \cite{Carillo:Schiebold:JMP2009}, \cite{SIGMA2016},  
\cite{MATCOM2018}, 
is devoted to study nonlinear evolution equations in which the unknown is an operator 
on a Banach space. 
The idea of the method can be sketched as follows:
\begin{itemize}
\item[$\bullet$] Consider the operator method to obtain pKdV solutions.
\item[$\bullet$] Use the links, via B\"acklund transformations,  between the  pKdV, KdV and mKdV equations to construct solutions of the KdV and  mKdV equations.
\item[$\bullet$] Implementation and visualisation of the explicit solutions via computer algebra routines. 
\end{itemize}
The quite involved technical details can be found in \cite{Carillo:Schiebold:JMP2011}.
The following first two subsections provide the schematic idea of the adopted method. In the third subsection
some explicit soliton solutions of the $2\times2$ and $3\times3$-matrix mKdV equation are presented.

\subsection{Operator method}

The solution method, \cite{AdenCarl}, \cite{Schiebold1}, can be sketched via the following diagram
{\scriptsize \begin{center}
\unitlength0.57cm
\begin{picture}(14,7.5)
   \put(-0.5,6){\framebox(4.5,1.1){\shortstack{{\it original} \\ soliton equation}}}
   \put(-0.5,5){\framebox(4.5,1){\shortstack{ solution \hspace*{\fill}
                                            \\ $u=u(x,t;a)$, $a\in\mathbb{C}$}}}
   \put(-0.5,1){\framebox(4.5,2.25){\shortstack{ solution formula with  \\
                                               an operator-valued \\
                                               parameter}}}
   \put(-0.50,0.25){\framebox(4.5,0.75){$\hat{u} = \tau(U(x,t;A))$}}
   \put(9,6){\framebox(5.5,1.1){\shortstack{{\it operator-valued}
                                           \\ soliton equation}}}
   \put(9,5){\framebox(5.5,1){\shortstack{ solution \hspace*{\fill}
                                           \\ $U=U(x,t;A)$, $A\in{\cal L}(E)$}}}
   \put(5.25,6){\shortstack{appropriate\\translation}}
   \put(5.5,1.4){\shortstack{ scalarization \\[0.3cm] }}
   \thicklines
   \put(4.5,5.75){\vector(1,0){4}}
   \put(11.5,4.5){\line(0,-1){2}}\put(10.75,2.5){\oval(1.5,1.5)[br]}
                                 \put(10.75,1.75){\vector(-1,0){6.5}}
   \put(1.5,4.75){\line(0,-1){0.875}}\put(2,4.75){\line(0,-1){0.875}}
      \put(1.5,4.75){\line(1,0){0.5}}\put(1.75,3.75){\line(-2,1){0.75}}
      \put(1.75,3.75){\line(2,1){0.75}}
\end{picture}
\end{center}
}\noindent
where $a$ and $A$ are parameters which, in turn, represent a generally complex number and an operator.   

\subsection{Soliton solutions of mKdV matrix equation}
			
The following theorem (cf. Theorem 11 b) in \cite{Carillo:Schiebold:JMP2011} where the matrix mKdV 
hierarchy is treated) allows to construct noncommutative soliton solutions of the mKdV equation
\begin{equation} \label{matrix mkdv}
	V_t = V_{xxx} + 3 \{ V^2, V_x \} .
\end{equation} 
More precisely,  \eqref{matrix mkdv} is to be read as an equation in which the dependent 
variable $V=V(x,t)$ takes values in the ${\sf d}\times{\sf d}$-matrices.
For the proof and the notation related to Banach spaces we refer to \cite{Carillo:Schiebold:JMP2011}.

\begin{theorem} \label{T3 KdV mKdV}
   	Let $E$ be a Banach space and $\cal A$ a quasi-Banach ideal equipped with
   	a continuous determinant $\delta$. Let $A\in{\cal L}(E)$ with $\mbox{\rm spec}(A)
   	\subseteq \{ \lambda\in\mathbb{C} | \mbox{\rm Re}(\lambda) > 0 \}$.
   	Moreover, assume that $B\in{\cal A}(E)$ satisfies the ${\textsf{d}}$-dimensionality condition
   	\begin{equation} \label{d dimensionality}
      		AB+BA = \sum_{j=1}^{\textsf{d}} d^{(j)}\otimes c^{(j)}
   	\end{equation}
   	with linearly independent $d^{(j)}\in E'$ and $c^{(j)}\in E$, $j=1,\ldots,{\textsf{d}}$.
	Then the matrix-function  \vskip0em \noindent
	\begin{equation*}
         	V = \frac{\rm i}{2} \
             		\biggm(
               \frac{\delta\bigm(I+\mbox{\rm i}(L+L^{(i,j)})\bigm)}{\delta\bigm(I+\mbox{\rm i}L\bigm)}
                  - \frac{\delta\bigm(I-\mbox{\rm i}(L+L^{(i,j)})\bigm)}{\delta\bigm(I-\mbox{\rm i}L\bigm)}
             	\biggm)_{i,j=1}^{\textsf{d}} ,
      	\end{equation*}\vskip-1em \noindent
	where
	\begin{equation}
      		L^{(i,j)} = d^{(i)}\otimes \Big(\exp\big( Ax+A^3t \big)\,c^{(j)}\Big), \quad
      		L=\exp \big( Ax+A^3t \big) B,
   	\end{equation}
   	and $I=I_E$ denotes the identity operator on $E$,
	solves the matrix mKdV \eqref{matrix mkdv} with values in the 
	${\sf d}\times{\sf d}$-matrices
	on every product domain $(-\infty,c)\times G$ ($c\in\mathbb{R}$,
      	$G$ a domain in $\mathbb{R}^{n-1}$) on which $\delta(I\pm \mbox{\rm i}L)\not=0$.
\end{theorem}
\vskip-5em
\subsection{Some explicit solutions of the matrix modified KdV equation}%
The final step consists in the implementation, via computer algebra, of suitable routines amenable to visualise the matrix solutions of the mKdV equation. Indeed, multisoliton solutions admitted by the matrix KdV equation are given by Goncharenko in  \cite{Goncharenko}: these solutions are obtained via a generalisation of the Inverse Scattering Method. As shown in \cite{Carillo:Schiebold:JMP2011}, the solution class of the matrix KdV equation that corresponds to the class constructed in Theorem \ref{T3 KdV mKdV} comprises Goncharenko's multisoliton solutions.
In the present subsection, soliton solutions of the matrix \emph{modified} KdV are visualised using Mathematica.
Some of such explicit 2-soliton solutions of the matrix mKdV, in the present subsection, 
are visualised using {\it Mathematica}.

Without going into details, we briefly explain the choices in Theorem \ref{T3 KdV mKdV} to realize 2-soliton solutions of the ${\sf d}\times {\sf d}$-matrix mKdV. To reflectionless spectral data with discrete 
eigenvalues $k_1,k_2$ and spectral matrices $B_1,B_2$ of size ${\sf d}\times {\sf d}$, we associate
    $ A =\begin{pmatrix} k_1 I_d & 0 \\ 0 & k_2 I_d \end{pmatrix}, \qquad
                          B = \Big( \frac{{\rm i}}{k_i+k_j} B_j \Big)_{i,j=1}^2,$
                          \noindent
In particular, $E=\mathbb{C}^{2\sf d}$.
\smallskip
In the plots below  the case $B_1=B_2$ is considered. 
\begin{figure}[H]
\centering
\vskip-2.5em
\epsfig{file=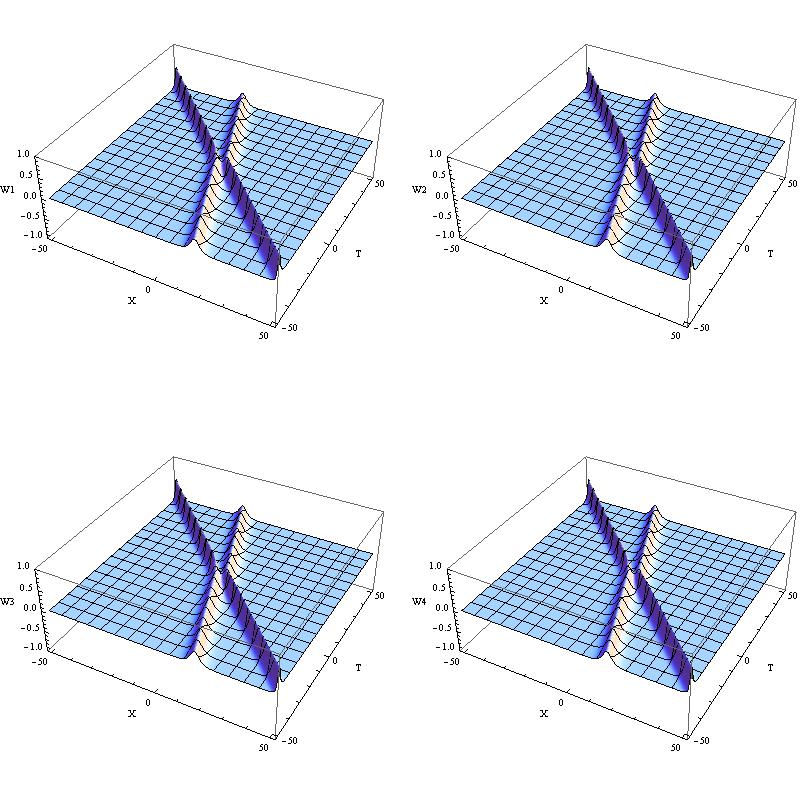, scale=1.65}\hskip.4cm\epsfig{file=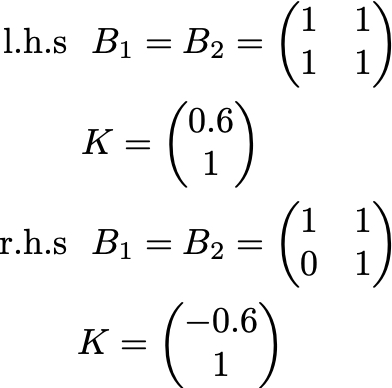, scale=0.22}\hskip.4cm\epsfig{file=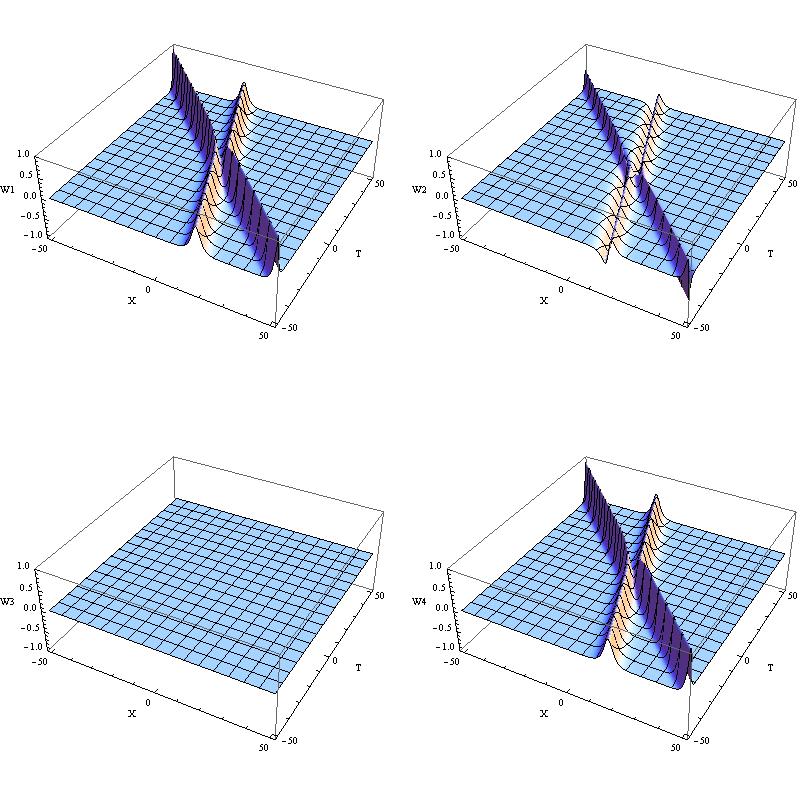, scale=1.65} \\

\vskip-1em
\caption{Two examples of two-soliton solutions  of   $2\times2$   matrix mKdV equation;  the  elements of the spectral matrices are specified.}%\vskip-3em
\label{fig4}
\end{figure}
\vskip-1em
The examples of solutions of the  mKdV equation, graphically represented in the Fig.s 2 - 4,  
represent different  $3 \times 3$ matrix solutions: 
the elements of the spectral matrices are indicated in the  caption. % some. 
The pictures show the behaviour of  {\it two-soliton solutions}  admitted by the matrix mKdV equation. 

\vskip-1.3em\begin{figure}[H]
\epsfig{file=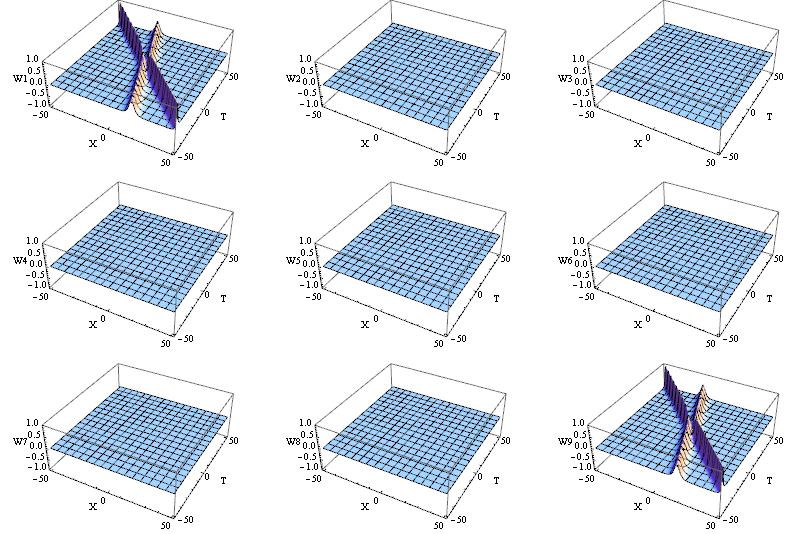, scale=0.25} \\

 \vskip-12.3em
 \hfill 
 \epsfig{file=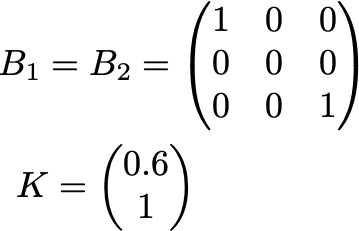, scale=0.30}  \vskip2.9em
\caption{Two-soliton solutions  of   $3\times3$   matrix mKdV equation;  the two diagonal elements $b_{1,1}=b_{3,3}=1$ all the others are equal to zero.}%\label{fig4}
\end{figure}
\vskip-3.2em
\vskip-1.1em
\begin{figure}[H]
\epsfig{file=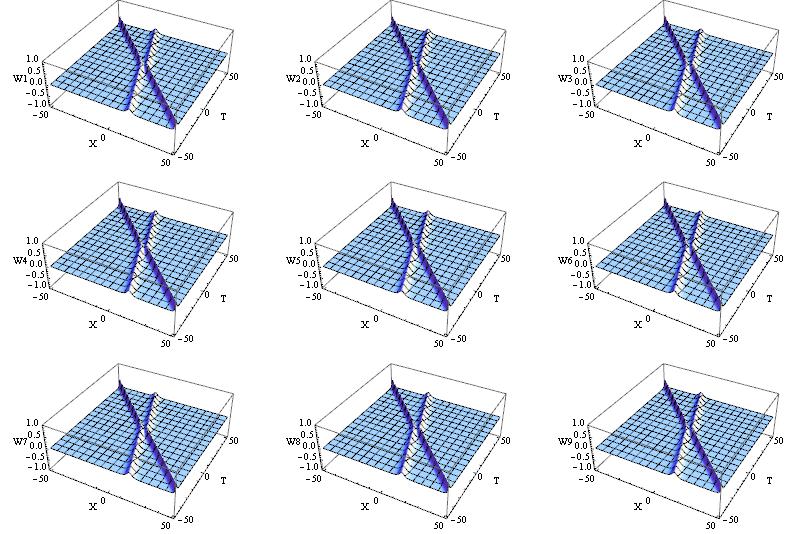, scale=0.25} \\
 \vskip-12.3em
 \hfill 
 \epsfig{file=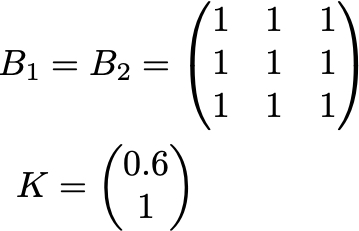, scale=0.30}  \vskip2.9em
  \caption{Two-soliton solution  of   $3\times3$   matrix mKdV equation; all the matrix elements $b_{h,k}=1$,
  $1\le h, k \le 3$.}
\label{fig5}\vskip-1.0em
\end{figure}
\begin{figure}[H]
\vskip-3.5em
\epsfig{file=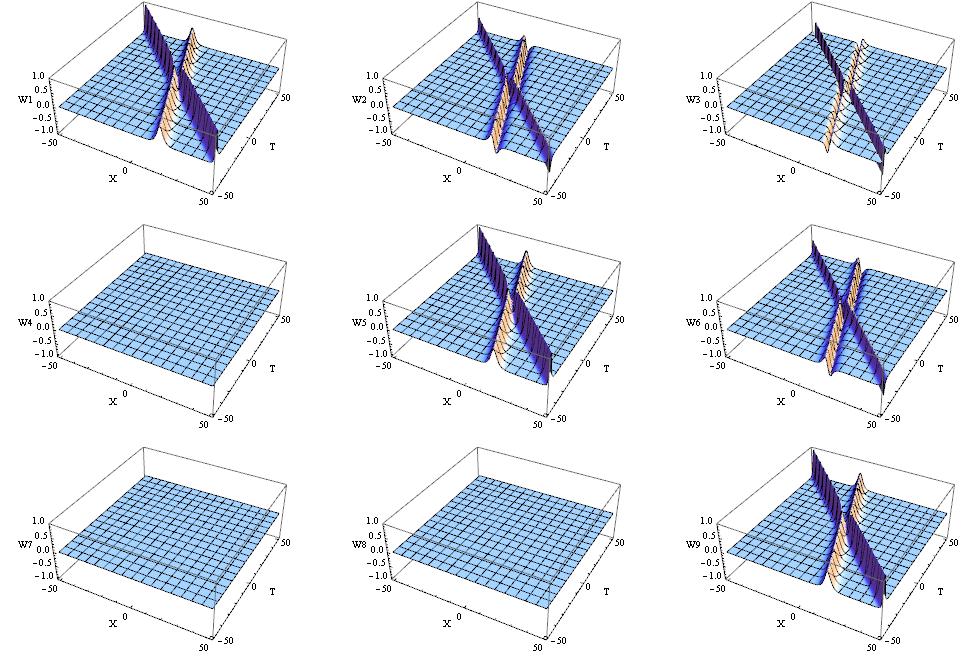, scale=0.20} \\

 \vskip-12.3em
 \hfill 
 \epsfig{file=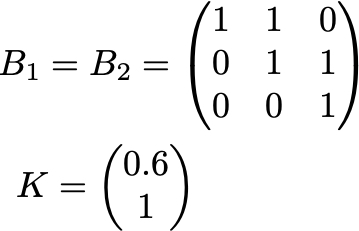, scale=0.30}\vskip2.3em
\caption{Two-soliton solution  of   $3\times3$   matrix mKdV equation;  all the matrix elements $b_{h,h}=1$,
  $1\le h \le 3$ and $b_{h,h+1}=1$,  $1\le h \le 2$  all the others are equal to zero.}
\vskip-1.5em\label{fig6}\vskip-.5em
\end{figure} 

Note that  the crucial features of interaction between two different solitons, 
well known on the scalar case seems to characterise also matrix solutions. 
These pictures represents only an 
example of a study, currently in progress, aiming to construct
additional solutions  as well as  their interpretation. In particular, if the  spectral 
matrices are diagonal, then also the solution enjoys the same property, see Fig. 2. Conversely, Fig. 
4, shows that, when  {\it off diagonal} elements in the spectral matrix are 
different from zero, the situation changes. For example, 
given the initial datum $b_{1,3}=0$, in the spectral matrices, the corresponding 
element of the two-soliton solution, depicted in Fig. 4, is not  zero.
A detailed study on these solutions  is currently under investigation. 
One of the main issues concerns the energy conservation and its partition among
the matrix elements. The appropriate functional which represents the energy of the 
interacting solitons is expected to play a crucial role to understand the 
phenomenology under investigation.   

\vskip-1em
\section{Concluding remarks}\label{4}
The aim of the present study is to emphasise  some of the properties of  solutions 
admitted by matrix  mKdV equation. Notably, depending on the spectral data, a
variety of soliton solutions may be observed. The crucial feature seem to be the 
appearance, also in the matrix case, of localised solutions which can be termed 
{\it solitons} on the basis of their interaction properties. 
A much richer interaction 
phenomenology, with respect to the scalar case,  can be observed when matrix 
 solutions are investigated. Indeed,
as soon as the spectral matrices have non-zero off-diagonal terms, the solution
exhibits non-zero solutions in further matrix elements. In addition,  a variety of 
different solutions can be observed. However, in all cases, a form of energy distribution
seems to be observed: the most appropriate way to define a functional suitable to 
represent the {\it energy} related to a matrix soliton solution is, in the authors 
opinion, one of the interesting questions this work arises.  
The obtained results motivate  further  investigations to provide a better 
understanding of the interesting phenomenology already observed.

\vskip-2em\subsection*{Acknowledgements}
Under the financial support of G.N.F.M.-I.N.d.A.M.,  I.N.F.N. and Universit\`a di Roma 
 \textsc{La Sapienza}, Rome, Italy.
C.Schiebold acknowledges  Dip. S.B.A.I., Universit\`a di Roma  \textsc{La Sapienza}, for the kind hospitality.
%%%%%%%%%%%%%%%%%%%%%%%%%%%%%%%%%%%%%%

\end{document}